\documentclass[letterpaper,longbibliography,twocolumn,showpacs,floatfix,groupaddress,twocolumns]{revtex4-1}
\usepackage[latin1]{inputenc}
\usepackage{bm}
\usepackage{multirow}
\usepackage{amssymb}
\usepackage{amsbsy}
\usepackage{amsmath,  amsthm, amsfonts, mathrsfs}
\usepackage{graphicx}
\usepackage{epsfig}
\usepackage{placeins}
\usepackage{wrapfig}
\usepackage{float}
\usepackage{trfsigns}
\usepackage[usenames,dvipsnames]{color}
\usepackage[colorlinks,linkcolor=blue,citecolor=blue,urlcolor=blue,breaklinks=true]{hyperref}
\newcommand{\mcl}{\mathcal{L}}
\newcommand{\mco}{\mathcal{O}}
\newcommand{\mch}{\mathcal{H}}
\newcommand{\mcq}{\mathcal{Q}}
\newcommand{\mcv}{\mathcal{V}}
\newcommand{\mce}{\mathcal{E}}
\newcommand{\mca}{\mathcal{A}}
\newcommand{\mcb}{\mathcal{B}}
\newcommand{\ii}{\mathrm{i}}

\newcommand{\reff}[1]{Ref.\ \cite{#1}}
\makeatletter

\begin{document} 

 \title{Numerically Probing the Universal Operator Growth Hypothesis}
 
 \author{Robin Heveling}
 \email{rheveling@uos.de}
 \affiliation{Department of Physics, University of Osnabr\"uck, D-49076 Osnabr\"uck, Germany}

 \author{Jiaozi Wang}
 \email{jiaowang@uos.de}
 \affiliation{Department of Physics, University of Osnabr\"uck, D-49076 Osnabr\"uck, Germany}

 \author{Jochen Gemmer}
 \email{jgemmer@uos.de}
 \affiliation{Department of Physics, University of Osnabr\"uck, D-49076 Osnabr\"uck, Germany}

\begin{abstract}
Recently, a hypothesis on the complexity growth of unitarily evolving operators was presented. This hypothesis states that in generic, non-integrable many-body systems the so-called Lanczos coefficients associated with an autocorrelation function grow asymptotically linear, with a logarithmic correction in one-dimensional systems.
In contrast, the growth is expected to be slower in integrable or free models. In the paper at hand, we numerically test this hypothesis for a variety of exemplary systems, including 1d and 2d Ising models as well as 1d Heisenberg models. While we find the hypothesis to be practically fulfilled for all considered Ising models, the onset of the hypothesized universal behavior could not be observed in the attainable numerical data for the Heisenberg model. The proposed linear bound on operator growth eventually stems from geometric arguments involving the locality of the Hamiltonian as well as the lattice configuration. 
We investigate such a geometric bound and find that it is not sharply achieved for any considered model.
\end{abstract}
\maketitle

\section{Introduction}
\label{int}
\noindent
The issue of the emergence of irreversible behavior from the unitary time evolution of quantum mechanics has yet to be answered in a satisfying manner \cite{gogolin}. In this context, concepts like the ``eigenstate thermalization hypothesis'' \cite{deutsch,sred,ales} and ``quantum typicality'' \cite{lloyd,gold,rei} have been introduced as possible fundamental mechanism behind an eventual equilibration of isolated quantum systems. The idea of typicality is that an overwhelming majority of pure states (at some energy) give rise to corresponding thermal expectation values. Thus, it is quite likely that over the course of time a pure state eventually ends up in the giant ``bubble'' of typical states, signaling an apparent equilibration of the system. In the Heisenberg picture formulation of quantum mechanics, not the states are time-dependent, but the observables themselves. Hence, it may be somewhat expected to find a similar notion of typicality for observables, going from initially simple, few-particle operators to more complex, generic operators. Recent works have studied this notion of operator growth from various angles \cite{op1,op2,op3,op4,op5,op6}.\\
\noindent
 In this paper, we refer to the particular work presented in Ref.\ \cite{berkley}, in which a hypothesis on the universality of operator growth is brought forth.
 Said hypothesis is formulated in the framework of the \textit{recursion method} \cite{recu2,recu} and makes a statement about the growth of the so-called Lanczos coefficients, real numbers that characterize the complexity growth of operators over the course of time. In the following, we numerically test this operator growth hypothesis for various models and observables. Further, we attempt to put the informal version of the hypothesis into a more quantitative context with regard to the functional form of the hypothesized universal growth pattern.

 
 \newpage
 
\noindent
The paper at hand is organized as follows:\ we briefly restate the universal operator growth hypothesis and introduce related quantities in Sec.\ \ref{ogh}. Following, in \mbox{Sec.\ \ref{mom}}, we derive an upper bound on the complexity growth of operators based on geometric arguments. We present our numerical results and relate them to the operator growth hypothesis in Sec.\ \ref{num}. We summarize our main results and conclude in Sec.\ \ref{conc}.

\section{Operator growth hypothesis}
\label{ogh}
\noindent
For self-containedness, in this section, we restate the operator growth hypothesis brought forward in Ref.\ \cite{berkley}. To start, the main quantities that eventually play a role in the hypothesis are introduced. We consider a system in the thermodynamic limit described by a local Hamiltonian $\mch$ [here, local means short-range, few-body interactions]. An observable of interest represented by a Hermitian operator $\mco$ gives rise to a corresponding autocorrelation function
\begin{equation}
\label{cor}
C(t) = \text{Tr}[\mco(t)\mco]\,,
\end{equation}
where $\mco(t)=e^{\mathrm{i}\mch t}\mco e^{-\mathrm{i}\mch t}$ is the time-dependent operator in the Heisenberg picture ($\hbar=1$). In the following, it is convenient to work directly in the Hilbert space of operators and denote its elements $\mco$ as states $|\mco)$. This Hilbert space of operators is equipped with an inner product $(\mco_1|\mco_2)= \text{Tr}[\mco_1^\dagger \mco_2 ]$, which induces a norm via $||\mco||=\sqrt{(\mco|\mco)}$. The Liouvillian superoperator is defined by $\mcl|\mco)= [\mch,\mco]$ and propagates a state $|\mco)$ in time such that the autocorrelation function may be written as $C(t)=(\mco|e^{\ii\mcl t}|\mco)$.\\ 
The \textit{Lanczos algorithm} can be employed to calculate a tridiagonal representation of the Liouvillian $\mcl$ in a (finite) subspace determined by some ``seed'' observable $\mco$. To start the iterative scheme detailed below, we take the normalized initial state $|\mco_0)=|\mco)$, i.e., $(\mco|\mco)=1$, and set $b_1=||\mcl \mco_0||$ as well as  $|\mco_1)=\mathcal{L}|\mco_0)/b_1$. Then we iteratively compute\\
\begin{align}
|\mcq_n)&=\mathcal{L}|\mco_{n-1})-b_{n-1}|\mco_{n-2})\,,\\\nonumber
b_n&=||\mcq_n||\,,\\\nonumber
|\mco_n)&=|\mcq_n)/b_n\,.
\end{align}
The tridiagonal representation of the Liouvillian in the \textit{Krylov basis} $\{|\mco_n)\}$ is then given by
\begin{equation}
\label{matrix}
L_{mn} =(\mco_m|\mcl|\mco_n)= \begin{pmatrix}
0 & b_1 \vphantom{\vdots} & 0 & ...\\
\vphantom{...}b_1 & 0\vphantom{\vdots} & b_2 & \\
0 & b_2 & 0 & \ddots\\
\vdots &  & \ddots & \ddots
\end{pmatrix}\,,
\end{equation}
where the \textit{Lanczos coefficients} $b_n$ are real, positive numbers output by the algorithm. They can be interpreted as hopping amplitudes in a tight-binding model and their iterative computation is an elementary part of the recursion method \cite{recu2,recu}.\\ Before the hypothesis itself is stated, we will briefly present the relation between the Lanczos coefficients $b_n$ and the autocorrelation function $C(t)$ or, respectively, its Fourier transform the spectral function
\begin{equation}
\Phi(\omega) = \int_{- \infty}^{\infty} e^{-\ii\omega t}\, C(t)\,\text{d}t\,.
\end{equation}
There exists a (non-linear) one-to-one map between the Lanczos coefficients $b_n$ and the spectral function $\Phi(\omega)$, thus, a set of $b_n$'s uniquely determines $\Phi(\omega)$ and vice versa.
It can be shown that the Lanczos coefficients $b_n$ appear in the continued fraction expansion of $\Phi(\omega)$, i.e.,
\begin{equation}
\Phi(\omega) = \text{Re } \dfrac{1}{\ii \omega + \dfrac{b_1^2}{\ii\omega+\dfrac{b_2^2}{\ii\omega+...}}}\,.
\end{equation}

\noindent
The universal operator growth hypothesis brought forward in \reff{berkley} concerns the asymptotic behavior of the Lanczos coefficients $b_n$. The hypothesis can informally be stated as follows: The Lanczos coefficients $b_n$ should ``grow as fast as possible'' in non-integrable systems. It turns out that [as detailed below] the fastest possible growth rate is (asymptotically) linear, i.e.,
\begin{equation}
b_n \sim \alpha n + \gamma +o(1)
\end{equation}
for some real constants $\alpha >0$ and $\gamma$.  In the special case of a one-dimensional system, the fastest possible growth is sub-linear due to an additional logarithmic correction, i.e.,
\begin{equation}
b_n \sim A \,\dfrac{n}{\ln n} + o(n/\ln n)\,,
\end{equation}
where $A>0$ is a real constant and $o(g_n)$ denotes some real sequence $f_n$ with $\lim_{n \rightarrow \infty} |f_n/g_n|=0$.\\
\noindent
 These bounds on fastest possible (asymptotic) growth eventually originate from a powerful statement on the behavior of the spectral function $\Phi(\omega)$ for large $\omega$.
The spectral function usually features non-vanishing high-frequency tails for generic many-body systems. By means of geometric arguments these tails can be rigorously bounded by an exponential function such that
\begin{equation}
\label{specbound}
\Phi(\omega) \leq K e^{-\kappa |\omega|}\,,
\end{equation}
for some adequately chosen constant $K>0$ and decay constant $\kappa>0$, which is related to the geometry of the system \cite{deroeck}.
It can be shown that spectral functions actually featuring exponentially decaying tails give rise to asymptotically linear growth in the Lanczos coefficients \cite{Lubinsky,magnus}.
Therefore, the operator growth hypothesis is equivalent to an exponentially decaying spectral function and basically states that the Lanczos coefficients should grow as fast as ``permitted by the geometry''.
A few examples are known for which linear growth is analytically shown to be achieved \cite{berkley, cao}.


\section{Bound on growth via moments}
\label{mom}
\noindent
The asymptotically linear bound on the growth of the $b_n$ or, respectively, the exponential bound on the decay of the spectral function $\Phi(\omega)$ are ultimately a consequence of geometric arguments concerning the locality of the Hamiltonian and the observable as well as the specific lattice geometry of the system \cite{deroeck}. A straightforward way to apply these arguments is by considering the moments $\mu_{2n}$ of the autocorrelation function and determining an upper bound on these moments by taking  the respective geometry of the system into account. The moments of the autocorrelation function $C(t)$ are defined by
\begin{equation}
\label{mom1}
\mu_{2n} = \dfrac{\text{d}^{2n}}{\text{d}t^{2n}} C(t) \big{|}_{t=0}
\end{equation}
or, respectively, in terms of the spectral function
\begin{equation}
\mu_{2n} = \int\omega^{2n}\,\Phi(\omega) \,\text{d}\omega\,.
\end{equation}
Since $C(t)$ is an even function, all odd moments vanish. The information contained in the moments $\mu_{2n}$ is identical to the information conveyed by the Lanczos coefficients $b_n$. It is detailed in App.\ \ref{translate} how to translate between the two quantities. \\
Employing the Heisenberg equation of motion for time-dependent operators, Eq.\ \eqref{mom1} can be written as
\begin{equation}
\mu_{2n}=||\mcl^{n}\mco||^2\,.
\end{equation}
This quantity will be bounded from above in the following. We again consider a local operator $\mco$ with $||\mco||=1$ and a local Hamiltonian $\mch = \sum_\ell h_\ell$ with local terms $h_\ell$ or, respectively, a local Liouvillian $\mcl = \sum_k \ell_k$ with local terms $\ell_k = [h_k, \cdot\,]$. \\
The norm of a local Liovillian $\ell$ [we assume periodicity such that all $\ell_k$ are of the same type] applied to some operator $\mathcal{A}$ can be bounded by
\begin{equation}
\label{b1}
||\ell \mathcal{A}||  \leq \mce \,||\mathcal{A}||\,,
\end{equation}
where $\mce = E_\text{max} - E_\text{min}$ denotes the maximum eigenvalue of $\ell$ and $E_\text{max}$ ($E_\text{min}$) is the maximum (minimum) eigenvalue of the local Hamiltonian $h$.
 Equality holds if the operator $\mca$ is an eigenoperator of the local Liouvillian corresponding to the largest eigenvalue.
Utilizing the triangle inequality and iteratively applying \mbox{Eq.\ \eqref{b1}} yields
\begin{align}
\label{b2}
||\mcl^{n}\mco|| &= ||\sum_{k_1,...,k_n} \ell_{k_n}...\, \ell_{k_1} \mco||\nonumber\\
&\leq \sum_{k_1,...,k_n} ||\ell_{k_n}...\, \ell_{k_1} \mco||\nonumber\\ &\leq  \sum_{k_1,...,k_n}  \mce^n =  \mce^n N_\text{sum}(n)\,,
\end{align}
\noindent
where $N_\text{sum}$ denotes the number of terms in the sum [it is specified below which terms are actually counted]. This number typically grows quite fast with $n$ and can be exactly determined for simple geometries, as is presented at the end of this section. Consequently, the moments can be bounded as
\begin{equation}
\label{bound}
\mu_{2n}=||\mcl^{n}\mco||^2 \leq  \mce^{2n} N_\text{sum}^2(n)\,.
\end{equation}
\noindent
This bound on the moments is sharp, meaning first, no sequence of moments $\mu_{2n}$ calculated via Eq.\ \eqref{mom1} can possibly grow faster and, importantly, second, that this bound can in principle be actually achieved with an equality sign. This is the case if and only if two conditions are met. Firstly, to get an equality in the triangle inequality, all operators occurring in the sum must be collinear. Secondly, the largest eigenvalue $\mce$ must be realized at each application of a local Liouvillian. Indeed, it would be quite surprising if this bound were to be achieved tightly for physical systems. On the other hand, it may not be as far off as one might initially guess, since multiple applications of the same operator [here $\mcl$] to a state increase the overlap of the resulting state with the states at the edges of the spectrum [given that the initial state has some overlap with eigenstates corresponding to extremal eigenvalues]. \\
The fastest possible growing moments, i.e., moments equal to the r.h.s.\ of Eq.\ \eqref{bound}, can be translated to corresponding Lanczos coefficients as described in App.\ \ref{translate}. For later reference, we denote the coefficients obtained this way by $\mcb_n$. They depend on the energy scale $\mce$ as well as on the lattice geometry expressed through $N_\text{sum}$, both quantities are exactly known for the models studied below. Even though Eq.\ \eqref{bound} gives a rigorous upper bound on the moments, the $\mcb_n$ obtained from this bound do not necessarily constitute a pointwise upper bound on the $b_n$. In Sec.\ \ref{isi} we give more details on the interpretation of the coefficients $\mcb_n$.\\
\noindent
To end this section, we determine the number $N_\text{sum}$ for the simple case of a one-dimensional chain with nearest-neighbor interactions, i.e., $\mch=\sum_\ell h_{\ell,\ell+1}$. We start with a local operator $\mco$ whose support is only on site zero [the support of an operator contains all sites on which the operator is not equal to the identity, e.g., here $\mco=...\otimes I \otimes \mco_0 \otimes I \otimes ...$, where $I$ denotes the identity on a given site]. The operator $\mcl\mco$ consists of operators $\ell_{-1,0}\mco_0$ with support on sites $(-1,0)$ as well as $\ell_{0,1}\mco_0$ with support on sites $(0,1)$. Next, $\mcl^2\mco$ contains six operators with support on sites $(-2,-1,0)$, $(-1,0)$, $(-1,0,1)$, $(-1,0,1)$, $(0,1)$, $(0,1,2)$, respectively. In these lists we include [and count] trivially non-vanishing operators, i.e., operators that vanish due to a lack of overlap between respective supports are not counted [for example the operator $\ell_{8,9}\mco_0$ trivially vanishes], however, operators with respective overlap between supports [like $\ell_{0,1}\mco_0$] are always counted, even though the operator may vanish due to the specific choice of the local Hamiltonian and initial observable. In this manner, we iteratively apply the Liouvillian to the initial operator, grow the supports accordingly
and keep track of the number of potentially non-vanishing operators. For the case at hand, this procedure gives rise to a sequence $N_\text{sum} (n)=1,2,6,22,94,454,2430,14214,...$ of numbers of terms in the sum in Eq.\ \eqref{bound} [for this exact sequence there happens to exist a closed form:\ $N_\text{sum}(n)= \sum_{k=0}^n 2^k S_n^{(k)}$, where $ S_n^{(k)}$ denotes the Stirling number of the second kind].\\
Corresponding iteratively computed sequences can be found for more complicated local Hamiltonians, e.g., next-nearest neighbor interactions, quasi one-dimensional systems and two-dimensional systems.

\section{Numerical Analysis}
\label{num}
\noindent
In this section, we numerically check the proposed operator growth hypothesis by explicitly calculating the Lanczos coefficients $b_n$ for various exemplary setups. These setups include one-dimensional and two-dimensional Ising models as well as one-dimensional Heisenberg models, all paired with a variety of different observables. We compare the calculated $b_n$ with the coefficients $\mcb_n$ [obtained from the r.h.s.\ of Eq.\ \eqref{bound}] by explicitly calculating $\mce$ and $N_\text{sum}$ for each model.\\
In practice, it is only possible [for the considered systems at least] to obtain a finite number $N$ of Lanczos coefficients $b_n$, since the dimension of the operator space grows exponentially. The achievable number of coefficients $N$ is around $30$ for the $1$d Ising model and around $15$ for the Heisenberg model. The difference in obtainable $b_n$ is due to the fact that the operator space grows much faster for the Heisenberg model than for the Ising model. For all considered systems the Hamiltonian $\mch$ consist of two terms, an integrable part $\mch_0$ and an integrability-breaking part $\mcv$, i.e.,
\begin{equation}
\label{hami}
\mch=\mch_0 + \lambda \mcv \,
\end{equation}
\newpage

\vspace*{-6px}
\onecolumngrid
        \begin{figure}[t]
          \hypertarget{fig:test}{}
  \includegraphics[width=2.15\linewidth]{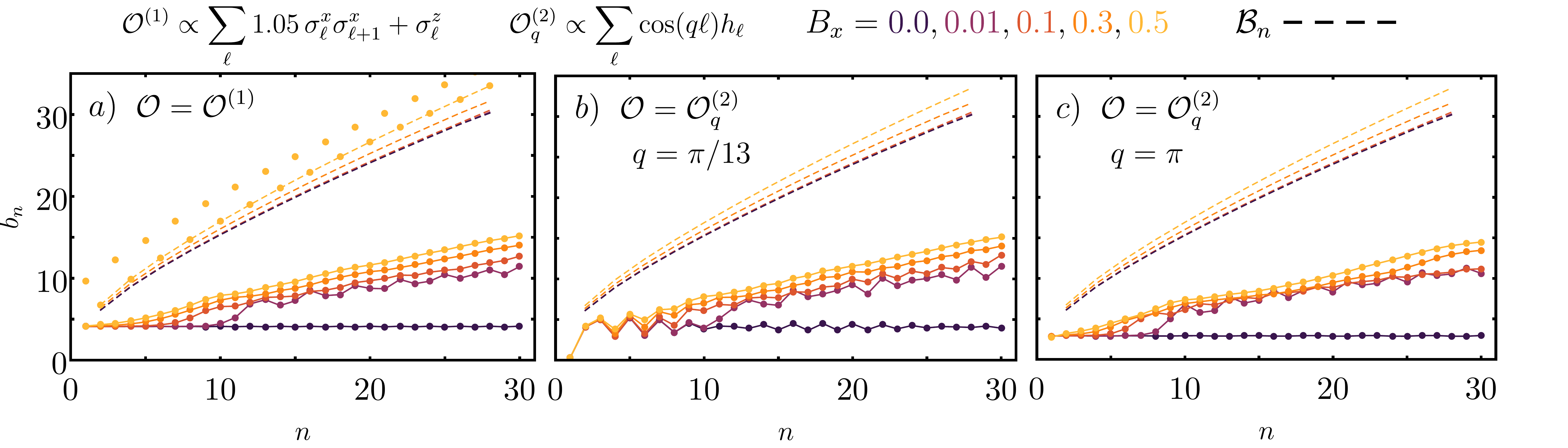}
  \end{figure}
\vspace*{-26px}
\noindent
FIG.\ 1.\ \vspace*{-2px}  Lanczos coefficients $b_n$ of the transverse Ising model for the $2$-local observables $a$) $\mco^{(1)}$, $b$) slow mode of $\mco^{(2)}_q$ with $q=\pi/13$, and $c$) fast mode of $\mco^{(2)}_q$ with $q=\pi$. The integrability-breaking magnetic field attains values from $B_x=0.0$ to $B_x=0.5$. For all observables, the transition from a free model to a non-integrable model is evident. The coefficients $\mcb_n$ are explicitly depicted in $a$) for the case $B_x=0.5$ as yellow dots. Dashed lines indicate the ``lower branches'' of the corresponding $\mcb_n$. To avoid clutter, only the dashed lines are depicted as a guide to the eye for other values of $B_x$ and in all following figures. The coefficients $\mcb_n$ are larger than the physical $b_n$ by a factor of about two for all observables.\\
\twocolumngrid
\setcounter{figure}{1}

\noindent
[except for the 2d Ising model, where $\mch_0$ is already non-integrable]. The parameter $\lambda$ tunes the non-integrability of the model. We suppose that the total Hamiltonian $\mch$ (as well as $\mch_0$ and $\mcv$ individually) can be written in terms of local Hamiltonians, i.e., $\mch=\sum_\ell h_\ell$. Again, the local terms usually describe short-range, few-body interactions.
For each model, we consider a number of observables $\mco$. Importantly, all observables should have zero overlap with any conserved quantity \cite{mazur}, for example \mbox{$(\mco|\mch)=0$}.

\subsection{Transverse Ising model}
\label{isi}
\noindent
The first model under consideration is a transverse Ising model with a tilted field. Respective unperturbed and total Hamiltonians are given by
\begin{align}
\mch_0= &\sum_\ell  J_{xx} \sigma^x_\ell \sigma^x_{\ell+1} + B_z \sigma^z_\ell\\ \mch&=\mch_0 + B_x \sum_\ell\sigma^x_\ell\,,\nonumber
\end{align}
where $\sigma_\ell^{x,y,z}$ denote Pauli operators on site $\ell$.
The magnetic field $B_x$ in $x$-direction plays the role of the integrability-breaking parameter $\lambda$ in Eq.\ \eqref{hami}, i.e., the system is non-integrable for $B_x \neq 0$ and integrable for $B_x=0$. We set $J_{xx} = 1.0$ and $B_z=-1.05$ and calculate the Lanczos coefficients for various observables as detailed in \mbox{Sec.\ \ref{ogh}}. In practice, it is convenient to adopt the set of Pauli strings as a working basis of the Hilbert space of operators \cite{berkley,pauli}.
As a first example, we consider the $2$-local observable
\begin{equation}
\label{obs1}
\mco^{(1)} \propto \sum_\ell 1.05 \,\sigma^x_\ell \sigma^x_{\ell+1} + \sigma^z_\ell\,,
\end{equation}
\newpage
\noindent
where $2$-local means that the local terms are supported on two sites respectively. The choice of the parameter $1.05$ in front of the $xx$-coupling term ensures that $(\mco^{(1)}|\mch)=0$. This exact setup was also studied in \mbox{Ref.\ \cite{berkley}}. The corresponding Lanczos coefficients are depicted in Fig.\ \hyperlink{fig:test}{1a} for various values of the integrability-breaking parameter $B_x$.\\
The Hamiltonian of the bare transverse Ising model [with $B_x=0$] can be mapped onto free fermions via a Jordan-Wigner transformation. Further, the observable  in \mbox{Eq.\ \eqref{obs1}} is local in the fermionic picture. In this non-interacting case, the Lanczos coefficients seem to be more or less constant. As soon as a small perturbation that breaks the integrability is introduced, e.g., $B_x=0.01$, the Lanczos coefficients begin to grow. The distinction between the free case and non-integrable cases is clearly visible in Fig.\ \hyperlink{fig:test}{1a}. The growth of the Lanczos coefficients for larger values of $B_x$ already looks quite linear, a possible logarithmic correction due to the system's one-dimensionality is not visible in the data. \\
  \noindent
Before we continue, we want to make some remarks on the interpretation of the coefficients $\mcb_n$. As mentioned, the $\mcb_n$ are computed by assuming that the moments grow as fast as possible, i.e., an equality sign in \mbox{Eq.\ \eqref{bound}},  and translating these maximum moments to Lanczos coefficients as detailed in App.\ \ref{translate}. However, even though \mbox{Eq.\ \eqref{bound}} gives a rigorous upper bound on the moments, the resulting $\mcb_n$ do not necessarily constitute a strict pointwise upper bound on the $b_n$ [this is simply due to the way the Lanczos coefficients are calculated from the moments]. Thus, the $\mcb_n$ should not be thought of as such.
Rather, the $\mcb_n$ represent a sort of ``global uniform'' upper bound, meaning that it is impossible to further increase the value of one specific coefficient ``by hand'', without simultaneously decreasing the value of another one [or several other ones]. In this sense the $\mcb_n$ are the ``maximum'' coefficients.
If the ``physical'' coefficients $b_n$ would follow the $\mcb_n$ tightly, then one could indeed conclude that the upper bound in Eq.\ \eqref{bound} was sharply achieved and that the Lanczos coefficients would indeed grow ``as fast as possible'', given the constraints in Eqs.\ \eqref{b1}, \eqref{b2}.\\
\noindent
The values of the $\mcb_n$ are explicitly depicted in Fig.\ \hyperlink{fig:test}{1a} for $B_x=0.5$. The coefficients clearly exhibit some even-odd effects. These even-odd effects also occur for all other considered parameters and models. To avoid clutter, we only show the ``lower branches'' of the $\mcb_n$ as dashed lines for smaller values of $B_x$. Therefore, the dashed lines in Fig.\ \hyperlink{fig:test}{1a} [and all following figures] serve as a guide to the eye for the ``maximal possible'' coefficients $\mcb_n$. The energy scale is $\mce =3.2$ [for $B_x=0.5$] and $N_\text{sum}$ is obtained as detailed at the end of Sec.\ \ref{mom}, with the important difference that the initial operator is now supported on two sites. In Fig.\ \hyperlink{fig:test}{1a} it is evident that the coefficients $\mcb_n$ are larger than the physical coefficients $b_n$ by about a factor of $2$. Thus, the upper bound on the moments is not sharply achieved, i.e., Eq.\ \eqref{bound} is an overestimate \cite{note}.
\noindent
This general behavior of the $b_n$ as well as the $\mcb_n$ is reproduced for the next observable, which constitutes an energy density wave with momentum $q$, i.e.,
  \begin{equation}
  \label{obs2}
\mco_q^{(2)} \propto \sum_\ell \cos(q\ell) h_\ell\,.
\end{equation}

\begin{figure}[b]
\includegraphics[width=\linewidth]{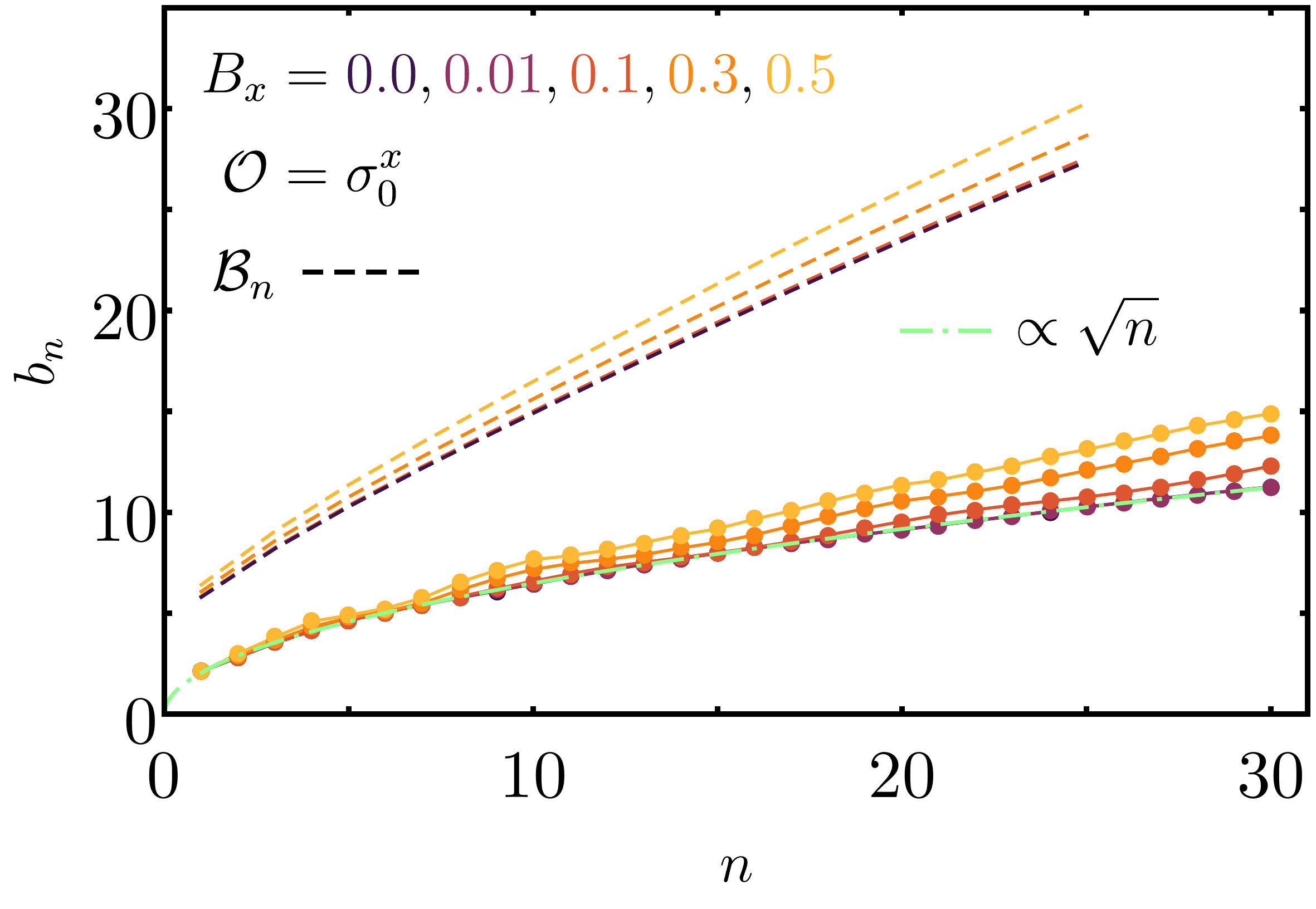}
  \caption{Lanczos coefficients $b_n$  of the transverse Ising model for a local observable $\mco^{(3)} = \sigma_0^x$ for various integrability-breaking parameters $B_x$. The distinction between the free and non-nonintegrable curves is not as striking a before. Green dash-dotted line indicates a fit $ \propto \sqrt{n}$ to the data of the integrable model. Dashed lines serve as a guide to the eye for the coefficients $\mcb_n$, which are larger by a factor of about two.}
  \label{isi_bn_loc}
  \end{figure}
\begin{figure}[t]
\includegraphics[width=\linewidth]{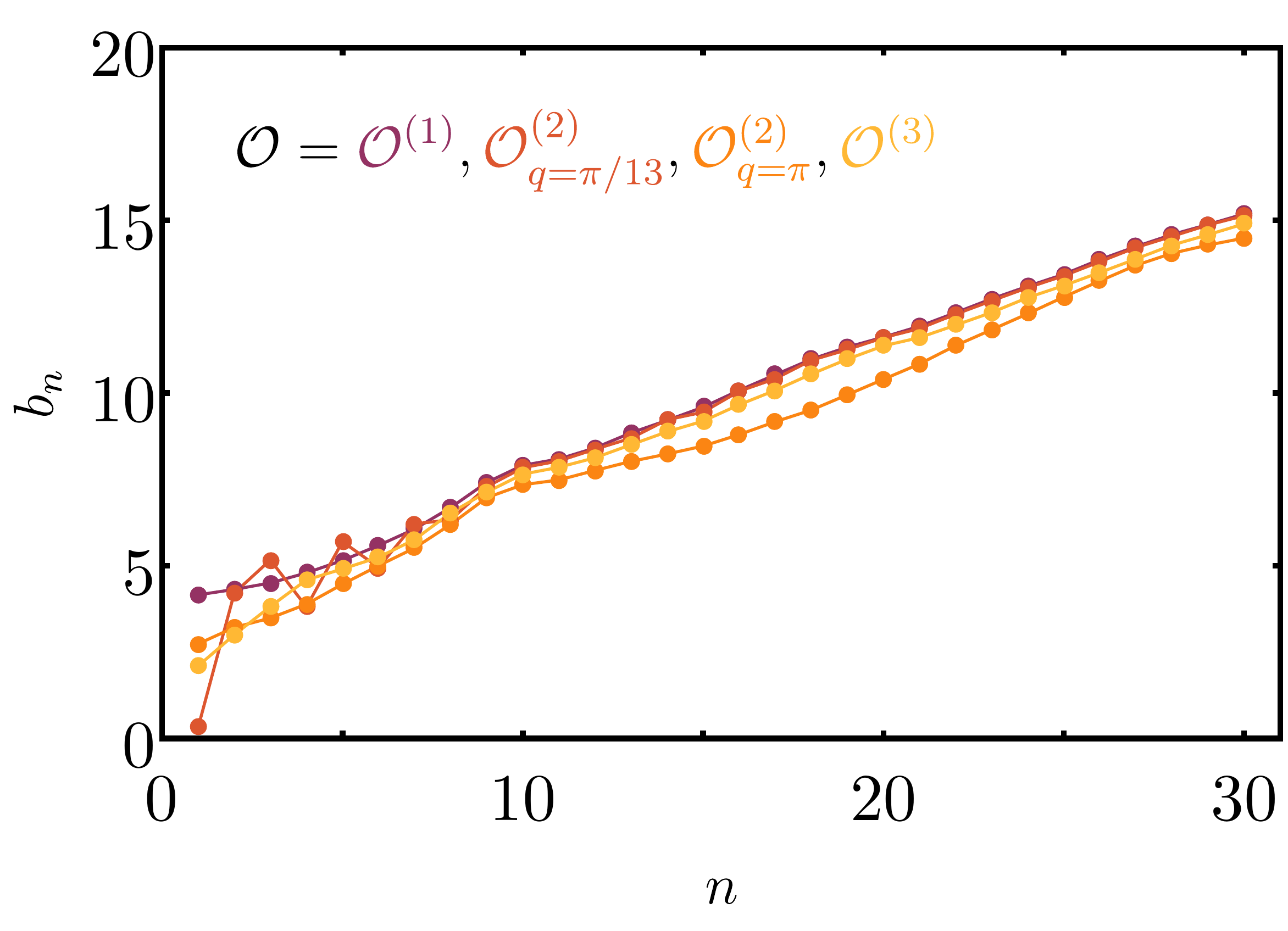}
  \caption{Comparison between Lanczos coefficients for the Ising model with $B_x=0.5$ for all four obervables considered thus far. The growth is quite similar for larger $n$ hinting at a universality of operator growth.}
  \label{isi_bn_comp}
  \end{figure}

 
\noindent
We study a relatively slow dynamic with $q=\pi/13$ and a relatively fast dynamic with $q=\pi$. Both observables are local in the fermionic picture.
The Lanczos coefficients $b_n$ and coefficients $\mcb_n$ are depicted in Fig.\ \hyperlink{fig:test}{1b} for $q=\pi/13$ and in Fig.\ \hyperlink{fig:test}{1c} for $q=\pi$.
In both cases, the qualitative behavior is quite similar to the first observable. Again, the Lanczos coefficients of the free model with $B_x=0$ seem to be more or less constant. Once the additional magnetic field is added, the model becomes non-integrable and at some point the $b_n$ grow approximately linearly. Just as for the first observable, the derived bounds are not tight and the $\mcb_n$ are larger by a factor of about two.\\
\noindent
The final considered observable for the one-dimensional Ising model is a local operator whose support only contains a single site, i.e.,
\begin{equation}
\label{obs3}
\mco^{(3)} = \sigma^x_0\,.
\end{equation}
The corresponding Lanczos coefficients are depicted in Fig.\ \ref{isi_bn_loc}. There is a clear qualitative difference compared to the observables investigated thus far. For the free case with $B_x=0$ the Lanczos coefficients seem to no longer be bounded by a constant. Rather, the growth is quite accurately described by a square-root $\propto \sqrt{n}$  [see fit]. For this particular model and observable the square-root-like growth can be understood analytically \cite{cao}. Further, this kind of growth has been observed for a variety of other integrable models \cite{recu,berkley,squ}. We suspect that this qualitatively different behavior compared to the previous cases is due to the specific choice of the observable, which is, in contrast to all previously considered observables, non-local in the fermionic picture. Consequently, one could be inclined to formulate two sufficient conditions, which both have to be met in order for the $b_n$ to be bounded by a constant. First, the Hamiltonian has to describe a free model and, second, the observable has to be local. For the observable at hand, which is non-local in the fermionic picture, the second condition is violated. Therefore, the Lanczos coefficients are not bounded by a constant, rather they grow as a square-root.\\
\noindent
As the Hamiltonian departs from the integrable/free point once $B_x\neq 0$, the $b_n$ grow faster [which is not too surprising, since there are simply more terms in the Hamiltonian]. From the computed data it is not immediately obvious whether the growth becomes linear (with a logarithmic correction) or remains more or less square-root-like. Possibly, the data for larger $B_x$ in Fig.\ \ref{isi_bn_loc} hints at an onset of linear growth for larger $n$. However, the distinction is certainly not as clear as for the local observables. Without previous knowledge of which coefficients belong to which $B_x$, it would be quite difficult to say if a set of $b_n$ belongs to an integrable/free or non-integrable model [barring the absolute values]. Therefore, calculating the $b_n$ as a potential method to determine or define \mbox{(non-)integrability} does not seem feasible, since the number at which universal behavior sets in may be larger than the practically computable number of Lanczos coefficients. Again, the $\mcb_n$ are off by a factor of about $2$. \\
\noindent
It is interesting to note that in the non-integrable models all considered observables seem to more or less lead to similar growth patterns and attain similar values for larger $n$. For comparison,  Fig.\ \ref{isi_bn_comp} depicts the Lanczos coefficients for all four observables considered thus far for $B_x=0.5$. This figure certainly supports the hypothesis of a universality of operator growth brought forth in Ref.\ \cite{berkley}. Particularly striking is the relation between the observable $\mco^{(1)}$ and the slow Fourier mode $\mco_{q=\pi/13}^{(2)}$ since for $n\gtrsim 10$ the coefficients practically coincide.

\begin{figure}[t]
  \includegraphics[width=\linewidth]{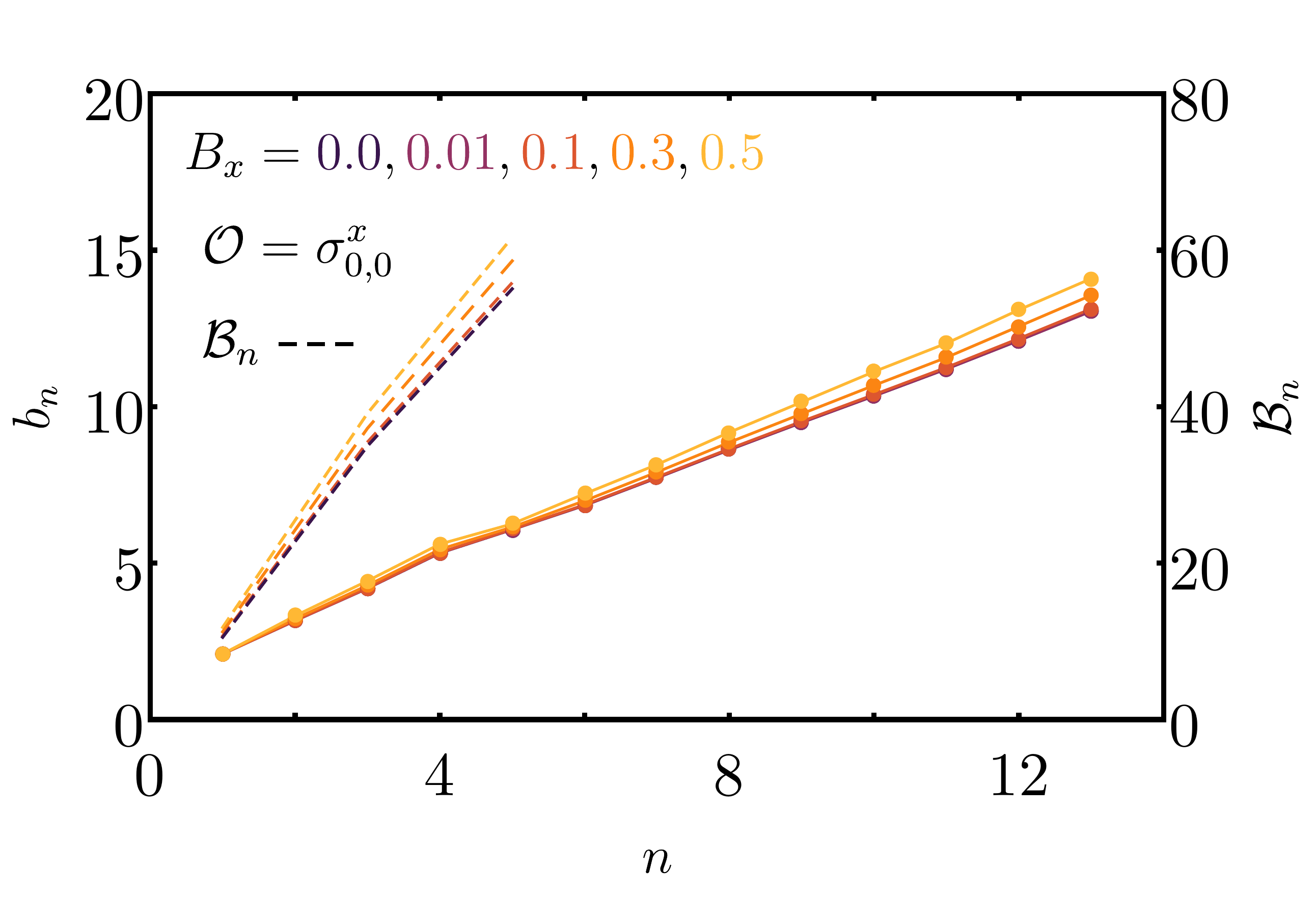}
  \caption{Lanczos coefficients $b_n$ of the two-dimensional transverse Ising model for an observable $\mco = \sigma_{0,0}^x$ for various $B_x$.
For all values of $B_x$ the growth is nicely linear.
Dashed lines serve as a guide to eye for the coefficients $\mcb_n$, which are much larger (note the additional vertical axis on the right).}
  \label{isi_bn_2d}
  \end{figure}

\noindent
Before leaving the Ising model and continuing with the Heisenberg model, we want to briefly present data on the  2d Ising model. As the two-dimensional Ising model is non-integrable, the hypothesis predicts a strict asymptotically linear  growth [without logarithmic correction] of the Lanczos coefficients $b_n$. Respective Hamiltonians of the two-dimensional Ising model are given by
  \begin{align}
\mch_0= \sum_{\ell,\ell'}  J_{xx}&\sigma^x_{\ell,\ell'} \sigma^x_{\ell +1 ,\ell'} + J_{xx}' \sigma^x_{\ell,\ell'} \sigma^x_{\ell ,\ell'+1} + B_z \sigma^z_{\ell,\ell'}\nonumber\\ &\mch=  \mch_0 + B_x \sum_{\ell,\ell'}\sigma^x_{\ell,\ell'}\,,
  \end{align}
where primed indices number the vertical direction and unprimed indices the horizontal direction.
\noindent
As in the one-dimensional case, we set $J_{xx}=1.0$, $B_z=-1.05$ and vary $B_x$. The coupling strength in vertical direction is set to $J'_{xx}=0.5$. 
Note that this model in non-integrable for all values of $B_x$. The energy scale is given by $\mce=3.9$ [for $B_x=0.5$]. We again consider a local observable whose support is only on one site, i.e.,
\begin{equation}
\mco=\sigma_{0,0}^x\,.
\end{equation}
The calculated Lanczos coefficients $b_n$ are depicted in \mbox{Fig.\ \ref{isi_bn_2d}}. Since the space of operators grows extremely fast, it is only practically possible to calculate about $13$ coefficients. These coefficients grow in a nicely linear fashion for all values of $B_x$, which is in accord with the operator growth hypothesis. The coefficients $\mcb_n$ from the derived upper bound on the moments are very far off. So far off, in fact, that there is an additional vertical axis for the coefficients $\mathcal{B}_n$ on the right side in Fig.\ \ref{isi_bn_2d}, which includes a factor of four. As mentioned, the number $N_\text{sum}$ is computed iteratively. For the two-dimensional Ising model the operator space grows so fast that the $\mcb_n$ are only attainable up to $n=5$.\newpage
\noindent
Summarizing the results from this section, the operator growth hypothesis is supported by (most of) the numerical data. The Lanczos coefficients of the one-dimensional non-integrable Ising models seem to eventually attain approximate linear growth for observables that are local in the fermionic picture. In these cases, the transition from free to non-integrable is clearly visible. A possible logarithmic correction is not noticeable in the presented data. Such a correction would probably only be visible in data that spans multiple orders of magnitude, which is not the case here. Further, the data for the two-dimensional Ising model supports the hypothesis for all considered values of $B_x$. Only the data for the third observable $\mco^{(3)}=\sigma_0^x$ is somewhat inconclusive. The transition is not as distinct as for the other observables, however, the onset of the hypothesized universal behavior may be suspected for larger $n$. All these numerical results support the in \mbox{Ref.\ \cite{berkley}} proposed operator growth hypothesis in the sense that the $b_n$ grow asymptotically linear. This is the first main result of the paper at hand.\\
In Ref.\ \cite{berkley}, the operator growth hypothesis is several times informally stated as:\ the Lanczos coefficients should ``grow as fast as possible'' and therefore eventually attain said linear growth. In fact, the data presented in this section corroborates the notion of asymptotically linear growth. However, the Lanczos coefficients could, in principle, grow much faster, as indicated by the coefficients $\mcb_n$, which are not tightly achieved in any of the considered models. Thus, we want to clarify the ``fastest possible growth'' is not to be understood with respect to the absolute numerical values of the coefficients, but rather in regard to the functional form of their growth. Indeed, the coefficients $b_n$ seem to grow with quite a similar functional form
as the [much to large] coefficients $\mcb_n$. This is the second main result of the paper at hand.

  \newpage
  
  \onecolumngrid
      \begin{figure}[t]
      \centering
      \hypertarget{fig:test2}{}
  \vspace*{-35px}
  \hspace*{30px}
  \includegraphics[width=1.8\linewidth]{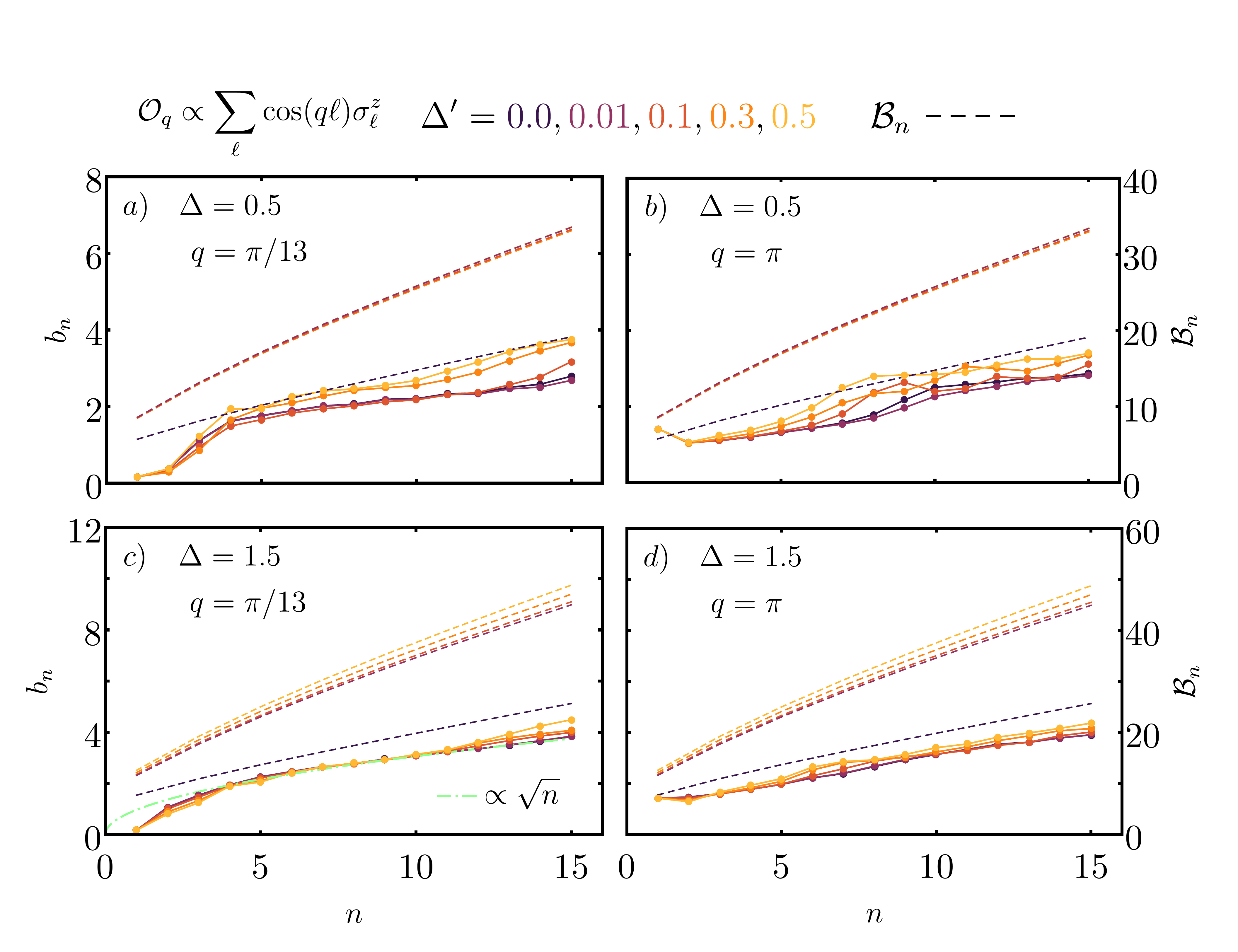}
  \label{heiseall}
  \end{figure}
\vspace*{-35px}
\hspace*{-14px} FIG.\ 5.\  Lanczos coefficients $b_n$ of the Heisenberg model for spin density waves. Depicted are various combinations of anisotropies and momenta:\ $a$) $\Delta=0.5$, $q=\pi/13$; $b$) $\Delta=0.5$, $q=\pi$; $c$) $\Delta=1.5$, $q=\pi/13$; $d$) $\Delta=1.5$, $q=\pi$. The integrability-breaking parameter $\Delta'$ attains values from $0.0$ to $0.5$. Dashed lines serve as a guide to the eye for the coefficients $\mcb_n$, which are much larger (note the additional vertical axes on the right).\\
\vspace*{25px}

\twocolumngrid
\setcounter{figure}{5}
\subsection{Heisenberg model}
\label{heisi}
\noindent
The second model of interest is a Heisenberg model with an additional next-nearest-neighbor interaction. Respective Hamiltonians are given by
\begin{align}
\label{heisimo}
\mch_0 =\sum_\ell &\sigma^x_\ell  \sigma^x_{\ell+1} + \sigma^y_\ell \sigma^y_{\ell+1}  + \Delta \sigma^z_\ell \sigma^z_{\ell+1}\\
\mch &= \mch_0 +  \Delta' \sum_{\ell,\ell'}\sigma^z_\ell \sigma^z_{\ell+2} \,.\nonumber
\end{align}

\noindent
The anisotropy of the nearest-neighbor interaction is denoted by $\Delta$ and the integrability-breaking next-nearest-neighbor interaction is tuned by the parameter $\Delta'$ [which plays the role of $\lambda$ in Eq.\ \eqref{hami}]. The bare Heisenberg chain [with $\Delta'=0$] is integrable for any anisotropy $\Delta$.\\ The model is gapless and exhibits ballistic transport behavior [of spin and energy] for $|\Delta| < 1$, whereas for $|\Delta| > 1$ the transport of spin is diffusive, while energy transport is still ballistic \cite{diffu}. In the following numerics we cover both cases by choosing values $\Delta=0.5,1.5$. The parameter $\Delta'$ that breaks the integrability is varied in the same fashion as $B_x$ in the Ising model. Note that the full Hamiltonian in \mbox{Eq.\ \eqref{heisimo}} conserves the total magnetization in $z$-direction.
\noindent
Therefore, it is natural to consider a spin density wave with momentum $q$, i.e.,
\begin{equation}
\mco_q \propto \sum_\ell \cos(q\ell) \sigma^z_\ell\,.
\end{equation}
Similar as for the energy density wave in the Ising model, we study a relatively slow dynamic with $q=\pi/13$ [depicted in Figs.\ \hyperlink{fig:test2}{5a}, \hyperlink{fig:test2}{5c} for $\Delta=0.5,1.5$ respectively] and a relatively fast dynamic with $q=\pi$ [depicted in \mbox{Figs.\ \hyperlink{fig:test2}{5b}, \hyperlink{fig:test2}{5d}} for $\Delta=0.5,1.5$ respectively].\\
\noindent
Since the Heisenberg Hamiltonian contains more coupling terms than  the Ising Hamiltonian, the dimension of the operator space [with respect to the Pauli basis] grows faster and we are restricted to a smaller number of coefficients, only about $15$.\\ 
Comparing the variance (relative deviations) of the Lanczos coefficients for $\Delta=0.5$ and $\Delta=1.5$, it is striking that the coefficients for $\Delta=1.5$ vary much less for different values of $\Delta'$. This is most likely due to the relative strength of the perturbation. Let $||\mch_0^\Delta||$ denote the norm of the unperturbed Hamiltonian with anisotropy $\Delta$ and $||\lambda\mcv||$ the strength of the perturbation [where $\lambda$ conforms to $\Delta'$]. Then, for example \mbox{$||0.5\mcv||/||\mch_0^{1.5}||=0.34$} and \mbox{$||0.5\mcv||/||\mch_0^{0.5}||=0.47$} [for comparison, in the one-dimensional Ising model $||0.5\mcv||/||\mch_0||=0.28$]. Again, the stronger the perturbation, the faster the coefficients grow, which is simply due to additional terms in the Hamiltonian [in order to not obscure the main points, we refrain from rescaling the Hamiltonian accordingly]. Other than that, the growth is more irregular than in the Ising model, at least in the  regime where data is available. For both values of $\Delta$ the transition occurs between an integrable [$\Delta'=0$] and non-integrable [$\Delta'\neq 0$] model. However, for $\Delta=0.5$ there is neither square-root-like growth in the integrable case nor linear growth in the non-integrable case visible [for both modes with $q=\pi/13$ in \mbox{Fig.\ \hyperlink{fig:test2}{5a}} and $q=\pi$ in Fig.\ \hyperlink{fig:test2}{5b}]. For $\Delta=1.5$, $\Delta'=0.0$ and $q=\pi/13$ the growth of the coefficients is more similar to a square-root [see fit] and only visibly deviates for $n\gtrsim 12$, see Fig.\ \hyperlink{fig:test2}{5c}. For the faster mode with $q=\pi$ the growth seems more linear with relatively small deviations, see Fig.\ \hyperlink{fig:test2}{5d}.\\
The coefficients $\mcb_n$ are much larger than any of the $b_n$ such that the additional vertical axes contain a factor of five in all cases. The number of terms in the sum $N_\text{sum}$ grows quite a lot faster than in the Ising model due to the next-nearest-neighbor interaction. These terms must in principle be included as soon as $\Delta'$ attains an arbitrarily small value strictly greater than zero. This leads to a somewhat gross overestimation, since the energy scale $\mce$ remains basically unaltered for small $\Delta'$. This is visible in all figures for the Heisenberg model, where the coefficients $\mcb_n$ for $\Delta'=0$ are calculated with the smaller numbers $N_\text{sum}$ of nearest-neighbor interaction. In principle, one could improve the upper bound on the moments by introducing a  second energy scale $\mce_{\text{nnn}}$ of the next-nearest-neighbor interaction and count terms according to the appearance of nearest-neighbor terms $\ell_{k,k+1}$ and next-nearest-neighbor terms $\ell_{k,k+2}$. This is, however, more complicated and not in the spirit of the derivations presented in Refs.\ \cite{berkley,deroeck}.\\
Summarizing, the numerical data for all considered values of $\Delta$ and $q$ can neither really reject nor support the operator growth hypothesis [not least because data for larger $n$ is not available]. The transitions between integrable and non-integrable models are certainly less striking than for the Ising model. Again, only looking at the $b_n$ it would be impossible to say which coefficients belong to an integrable or non-integrable model. This possibly raises the question whether the distinction between integrability and non-integrability regarding the growth of the $b_n$ is an adequate distinction to make. As mentioned in Sec.\ \ref{ogh}, the Lanczos coefficients are uniquely determined by the autocorrelation function $C(t)$. The Heisenberg model for $\Delta=1.5$ and $\Delta'=0$ is integrable and exhibits diffusive transport behavior \cite{diffu}, which is usually attributed to chaotic, non-integrable systems. In view of this, it may be not too surprising that the operator growth hypothesis is not supported by the (limited) numerical data presented in this section. This is the third main result of the paper at hand.\newpage

\section{Conclusion}
\label{conc}
\noindent
The first main message of the paper at hand concerns the explicitly calculated numerical data on the Lanczos coefficients.
We numerically probed the universal operator growth hypothesis proposed in Ref.\ \cite{berkley}, which states that in generic, non-integrable systems the Lanczos coefficients grow asymptotically linear [with a logarithmic correction in 1d]. We explicitly calculated Lanczos coefficients $b_n$ for various combinations of models [including 1d and 2d Ising models as well as ballistic and diffusive Heisenberg models] and
observables [including energy and spin density waves as well as local observables]. We found that the Ising model data generally supports the operator growth hypothesis. In particular, as soon as an integrability-breaking perturbation is added to the Hamiltonian, the coefficients eventually attain a linear growth [this transition is more pronounced in the case of a free Hamiltonian with a local observable (in the fermionic picture) than in the case of a free Hamiltonian with a non-local observable (in the fermionic picture)]. Further, the two-dimensional Ising model exhibits clear linear growth. Inconclusive, however, remains the data for the Heisenberg model. For none of the combinations of considered parameters there is a clear distinction between the integrable and non-integrable cases. Of course, it may be possible that the hypothesized universal behavior only sets in at some larger $n$, which is not accessible by our numerical tools.\\
\noindent
The second main message of the paper at hand concerns the coefficients $\mcb_n$, which are obtained by considering the fastest growing moments and converting them into Lanczos coefficients. The informal version of the operator growth hypothesis is stated several times in Ref.\ \cite{berkley}, namely that ``the Lanczos coefficients should grow as fast as possible'' and a corresponding bound on the moments leading to linear growth [with a logarithmic correction in one dimension] is given \cite{note}. As seen in the available numerical data, even the  ``optimized'' bound in Eq.\ \eqref{bound} is not remotely tight and the ``physical'' Lanczos coefficients $b_n$ increase much slower than the ``fastest possible growing'' coefficients $\mcb_n$ in all considered models. Therefore, technically, the Lanczos coefficients do not grow as fast as possible. Nevertheless, the physical Lanczos coefficients $b_n$ seem to grow in a manner that is compatible with the ``functional form'' of the maximal growth, i.e., we observe more or less linear growth for the Ising models, only with a flatter slope than would be induced by the bound on the moments.
\section*{Acknowledgments}
\label{ack}
\noindent
We thank D.\ Parker for interesting discussion on this topic and for a comment on an earlier version of this paper. This work was supported by the Deutsche  Forschungsgemeinschaft  (DFG)  within  the Research Unit FOR 2692 under Grant No. 397107022 (GE 1657/3-2).
\bibliography{literature}

\appendix

\section{Translating between $b_n$ and $\mu_{2n}$}
\label{translate}
\noindent
For completeness, we briefly present the relation between the Lanczos coefficients $b_n$ and the moments $\mu_{2n}$ \cite{gray}.\\
\noindent
\textit{i. From moments to Lanczos coefficients:\\}
To calculate the Lanczos coefficients $b_n$ from a given set of moments $\mu_{2n}$ we proceed as follows: we define \mbox{$c_n = \mu_{2n}/\mu_0$} and compute determinants of certain matrices constructed from the normalized moments $c_n$, i.e.,
\begin{equation}
B_n = \det (c_{i+j})_{0\leq i,j \leq n-1}
\end{equation}
where $n \geq 2$ and $B_0=B_1=1$ as well as
\begin{equation}
C_n = \det (c_{i+j+1})_{0\leq i,j \leq n-1}
\end{equation}
where $n \geq 1$ and $C_0=1$. Then the Lanczos coefficients are obtained as fractions of determinants via
\begin{equation}
b_{2n}^2= \dfrac{B_{n+1}C_{n-1}}{B_nC_n} \,\,,\quad b_{2n-1}^2= \dfrac{B_{n-1}C_{n}}{B_nC_{n-1}}\,.
\end{equation}

\noindent
\textit{ii. From Lanczos coefficients to moments:\\}
We take the representation $L$ of the Liouvillian $\mcl$ in the Krylov subspace spanned by the vectors generated by the Lanczos algorithm, cf.\ Eq.\  \eqref{matrix}.
The moments $\mu_{2n}$ can be easily read off as the upper-left element of even powers $2n$ of the matrix $L$, i.e.,
\begin{equation}
\mu_{2n} = (L^{2n})_{00}\,.
\end{equation}

\end{document}